\documentclass{article}
\usepackage{spconf,amsmath,graphicx}

\usepackage{xcolor} 
\usepackage{hyperref} 
\usepackage{float} 
\usepackage{subcaption} 
\usepackage{booktabs} 
\usepackage{multirow} 
\usepackage{ragged2e}

\title{FRAGMENTVC: ANY-TO-ANY VOICE CONVERSION BY END-TO-END EXTRACTING AND FUSING FINE-GRAINED VOICE FRAGMENTS WITH ATTENTION}
\name{
    Yist Y. Lin$^*$ \quad
    Chung-Ming Chien$^*$ \quad
    Jheng-Hao Lin \quad
    Hung-yi Lee \quad
    Lin-shan Lee
    \thanks{$^*$ These authors contributed equally.}
}
\address{
    National Taiwan University\\
    College of Electrical Engineering and Computer Science\\
    \{r08922048, r08922080, r08922049, hungyilee\}@ntu.edu.tw, lslee@gate.sinica.edu.tw
}
\begin{document}
\ninept

\maketitle

\begin{abstract}
    Any-to-any voice conversion aims to convert the voice from and to any speakers even unseen during training, which is much more challenging compared to one-to-one or many-to-many tasks, but much more attractive in real-world scenarios.
    In this paper we proposed FragmentVC, in which the latent phonetic structure of the utterance from the source speaker is obtained from Wav2Vec 2.0, while the spectral features of the utterance(s) from the target speaker are obtained from log mel-spectrograms.
    By aligning the hidden structures of the two different feature spaces with a two-stage training process, FragmentVC is able to extract fine-grained voice fragments from the target speaker utterance(s) and fuse them into the desired utterance, all based on the attention mechanism of Transformer as verified with analysis on attention maps, and is accomplished end-to-end.
    This approach is trained with reconstruction loss only without any disentanglement considerations between content and speaker information and doesn't require parallel data.
    Objective evaluation based on speaker verification and subjective evaluation with MOS both showed that this approach outperformed SOTA approaches, such as AdaIN-VC and \textsc{AutoVC}.
\end{abstract}

\section{Introduction}
\label{sec:intro}

Voice conversion (VC) technologies are to convert the voice produced by a source speaker to sound like being produced by a target speaker.
Conventional approaches using Gaussian Mixture Models \cite{stylianou1998continuous} worked well, though were then outperformed by those based on Artificial Neural Networks (ANNs) \cite{desai2009voice}.
Traditionally, aligned parallel data uttering the same text by different speakers were needed in training, but those data were difficult to obtain.
In recent years many parallel-data-free ANN-based models were proposed \cite{chou2018multi-target}, including CycleGAN-VC \cite{kaneko2018cyclegan-vc} and StarGAN-VC \cite{kameoka2018stargan}.
But these models can only perform conversion among a predefined set of speakers.
On the other hand, any-to-any VC models aim to convert the voice from and to any speakers even unseen during training, which is much more challenging while much more attractive in real-world scenarios.

Here we present FragmentVC, a parallel-data-free ANN-based approach for any-to-any voice conversion with an encoder-decoder architecture.
FragmentVC uses the latent phonetic structure of the utterance of the source speaker obtained with Wav2Vec 2.0 \cite{baevski2020wav2vec} as the query to extract the fine-grained voice fragments in the utterance(s) of the target speaker and fuse them into the desired utterance, all based on the attention mechanism of Transformer \cite{vaswani2017attention} and achieved end-to-end.

As shown in the lower part of Fig.~\ref{fig:model_arch}, FragmentVC consists of a source encoder (left in red), a target encoder (middle in blue) and a decoder (right in purple) and is trained with a two-stage training process.
The source encoder relies on Wav2Vec 2.0 to obtain the latent phonetic structure of the utterance from the source speaker, and the target encoder extracts spectral features from the log mel-spectrograms of utterances from the target speaker.
Given the former as queries, the decoder learns to utilize the Transformer cross-attention to extract voice fragments from the utterance(s) of the target speaker and fuse them into the converted utterance.
FragmentVC is directly trained with reconstruction loss only without considering speaker-content disentanglement, but has been shown to generalize well on unseen speakers.
Further analysis showed that phonetically similar fragments of the utterances from the source speaker and the target speaker were implicitly aligned in the attention maps, implying that the attention mechanism actually achieved fine-grained unit-selection and concatenation,
the relatively coarse versions of which have long been used in text-to-speech \cite{donovan1998ibm, hunt1996unit} and VC \cite{takashima2012exemplar, jin2016cute}.

\section{Related Works}
\label{sec:related_works}

Concatenative text-to-speech \cite{donovan1998ibm, hunt1996unit} is to select voice segments from a corpus, concatenate them and smooth the boundaries.
A large set of transcribed utterances from a specific speaker is usually needed.
Some concatenative methods of VC were proposed earlier \cite{takashima2012exemplar, jin2016cute}.
They required parallel data and did one-to-one VC only.
But FragmentVC used the Transformer attention to achieve the purpose with voice fragments and perform any-to-any VC without parallel data.

Attention mechanism was used for VC in sequence-to-sequence (seq2seq) models \cite{tanaka2019atts2s-vc} or Transformer networks \cite{huang2019voice, liu2020voice} but both for one-to-one VC.
FragmentVC is not only any-to-any, not a seq2seq method, but also not using the attention to learn a monotonic alignment between the source and the converted utterances.
FragmentVC implicitly learns to use the latent phonetic information from the source utterance to extract and fuse the fine-grained voice fragments of the target utterances.

Any-to-any VC was achieved earlier in AdaIN-VC \cite{chou2019one-shot} and \textsc{AutoVC} \cite{qian2019autovc}.
They both relied on the disentanglement of content and speaker information within an utterance with a content encoder and a speaker encoder.
AdaIN-VC adopted instance normalization to convert from the source speaker to the target speaker, while \textsc{AutoVC} used a pretrained speaker encoder to obtain speaker information and used an information bottleneck to limit the leakage of the source speaker information.
Instead, FragmentVC directly attends on the utterances from the target speaker, extracting and fusing the proper fine-grained voice fragments, which was shown to perform comparably or better than AdaIN-VC and \textsc{AutoVC}.

\section{Methodology}
\label{sec:methodology}

\begin{figure}
    \centering
    \includegraphics{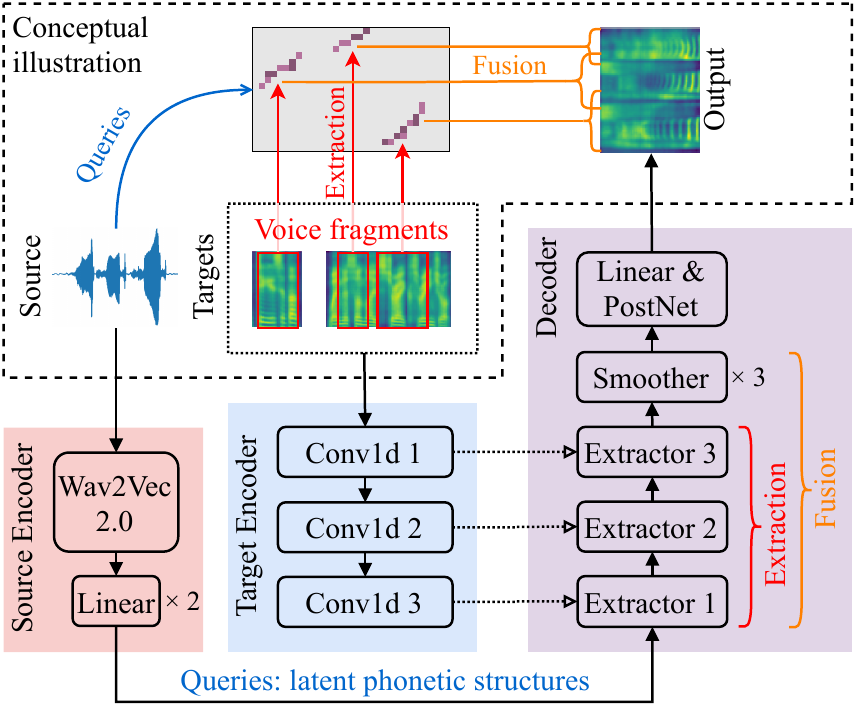}
    \vspace{-2em}
    \caption{
        The overall model architecture (lower half) and the concept of the process (higher half) of FragmentVC.
        The dotted arrows between the target encoder and the extractors indicate the attention.
    }
    \label{fig:model_arch}
\end{figure}

\begin{figure}
    \centering
    \includegraphics{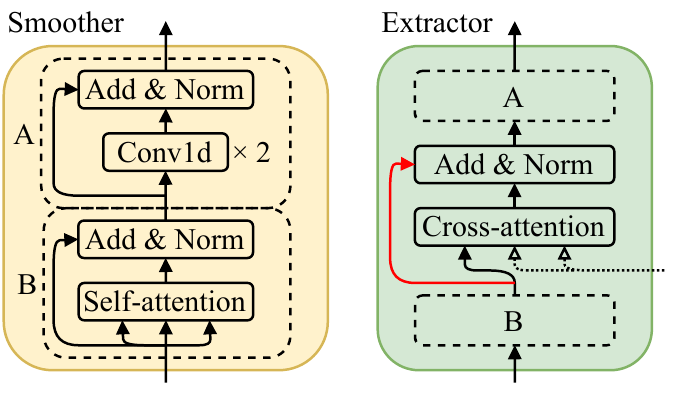}
    \vspace{-1em}
    \caption{
        The model architecture of the extractor and the smoother.
        Extractor~1 does not have the residual connection (red arrow).
    }
    \label{fig:ctl_arch}
\end{figure}

The overall model as shown in the lower part of Fig.~\ref{fig:model_arch} is composed of a source encoder, a target encoder, and a decoder.
The concept of the process is illustrated in the upper half of the figure.

\subsection{Source encoder}

Wav2Vec 2.0 \footnote{The pretrained model of Wav2Vec 2.0 Base trained on LibriSpeech \cite{panayotov2015librispeech} without finetuning on transcribed speech was used throughout this paper.} \cite{baevski2020wav2vec} is used as a pretrained feature extractor to extract 768-dimensional speech representations of the source utterance, with model weights fixed during training.
The 768-dimensional features are then converted to 512-dimension by two linear layers with ReLU activation, to be used as the input to the decoder.

\subsection{Target encoder}

The log mel-spectrograms of utterance(s) from the target speaker are concatenated and fed into the target encoder, which is composed of three ReLU-activated 1d-convolution layers, for extracting the voice fragments to be used below.

\subsection{Decoder}

The decoder is composed of a stack of extractors and smoothers, followed by a linear projection and a Tacotron-2-styled PostNet \cite{shen2018natural}, to predict the log mel-spectrogram for the desired output voice in a non-autoregressive manner.
Both the extractors and smoothers are Transformer \cite{vaswani2017attention} layers with two attention heads and hidden size being 512.
The extractors are equipped with both self-attention and cross-attention that attend on the output of the target encoder, while the smoother contains self-attention only.
Considering the high correlation among adjacent features in speech, the feed-forward layers in each Transformer layer are replaced by a convolutional network \cite{ren2020fastspeech}, with the detailed architecture shown in Fig.~\ref{fig:ctl_arch}.

The extractors are based on the latent phonetic structure of the source speaker utterance by cross-attention to extract fine-grained voice fragments from target speaker utterances, and then fuse them up to produce the output voice.
The cross-attention is purposely designed to have a U-Net-like \cite{ronneberger2015unet} architecture as in Fig.~\ref{fig:model_arch} as explained below.
Because Extractor~1 (lowest in Fig.~\ref{fig:model_arch}) is to construct the highest-level phonetic structure of the output utterance based on the source representations, while Conv1d~3 (also lowest in Fig.~\ref{fig:model_arch}) of the target encoder is supposed to produce the most abstractive spectral information too, so Extractor~1 attends on the output of Conv1d~3.
On the other hand, Extractor~3 (highest in Fig.~\ref{fig:model_arch}) is to offer only slight modifications or minor adjustments in the spectrogram, so it attends on the features obtained from Conv1d~1 (highest in Fig.~\ref{fig:model_arch}) of the target encoder.
The smoothers finally take the output of the extractor stack to further smooth the output utterances.

Since some residual speaker information is inevitably carried by the Wav2Vec features, we remove the residual connection over the cross-attention module in Extractor~1, as shown in red in Fig.~\ref{fig:ctl_arch}, to restrict the information flow from the source encoder to the decoder.
This was verified to be useful in the results in Sec.~\ref{sec:results}.

\subsection{Loss function}

Only L1 loss between the predicted and the ground-truth log mel-spectrograms is used to train the entire network, including the process of extracting and fusing the voice fragments, in an end-to-end manner, except for the fixed pretrained Wav2Vec model.
No additional loss term is needed as done previously \cite{chou2018multi-target, kaneko2018cyclegan-vc, kameoka2018stargan}.
FragmentVC was shown to obtain the target speaker characteristics with the target encoder.

\subsection{Two-stage training}
\label{ssec:two-stage}

We adopted a two-stage training scheme for the proposed model.
In the first stage, the same single utterance from a training speaker is used as the input of both the source encoder and the target encoder, and the training goal is to reconstruct the log mel-spectrogram of the utterance.
Although the spectrogram input of the target encoder is identical to the reconstruction target, there is no way for the attention mechanism to obtain the important absolute position information of the various acoustic events in the utterance from the spectral features.
So the model has to learn end-to-end to align the hidden structures between the Wav2Vec feature space from the source encoder and the spectral feature space from the target encoder by extracting and fusing the voice fragments.
Though the Wav2Vec features are not only very abstractive but also very different from the spectral features, we believe the linear layers in the source encoder provides some of the conversion between the two very different feature spaces.
Preliminary experiments showed the model was able to almost perfectly reconstruct the log mel-spectrogram in this way.
However, if a target utterance different from the source utterance was given at this stage, we found that the model was able to extract voice fragments properly and the output sounded like spoken by the target speaker, but the converted result sounded rather discontinuous because the model didn't learn to perform such a task.
This is why the second training stage as explained below is needed.

In the second stage, we concatenate the spectrograms of 10 utterances and feed them into the target encoder, while feeding a single utterance to the source encoder, all produced by the same speaker, with the training goal being reconstructing the spectrogram of the source utterance.
In the beginning, the source utterance (also the reconstruction target) is always included in the 10 target utterances, but the probability that it is included then linearly decays to zero as the training proceeds, so as to learn incrementally the scenario that the source and targets are getting more and more different.
In order to preserve the already well-trained attention modules, we reduce the learning rates of the source encoder, the target encoder and the extractors by 100 times in this stage, while the learning rates of the other components are left unchanged.

\begin{table*}
    \setlength{\tabcolsep}{2.5pt} 
    \centering
    \caption{
        The SV accuracy (\%) for seen-to-seen (with EER being $5.6\%$) and unseen-to-unseen (with EER being $2.6\%$) scenarios.
        10 target utterances were used except for columns (e) (f) (g).
    }
    \label{tab:speaker_verification_accuracies}
    \vspace{-0.75em}
    \begin{tabular}{cccccccccccc}
        \toprule
        \multirow{2}{*}{\textbf{Scenarios}} & \multicolumn{4}{c}{Comparison with other SOTAs} & \multicolumn{3}{c}{Different \# of target utterances} & \multicolumn{4}{c}{Ablation studies}                                                                                                                                                                                     \\
        \cmidrule(lr){2-5}
        \cmidrule(lr){6-8}
        \cmidrule(lr){9-12}
                                            & (a) $^\ast$Proposed                  & (b) $\ast$\textit{-ss}                                & (c) AdaIN-VC                         & (d) \textsc{AutoVC} & (e) 1 tgt & (f) 5 tgt & (g) 20 tgt & (h) $\ast$\textit{-ca} & (i) $\ast$\textit{-lrr} & (j) $\ast$\textit{-rrc} & (k) $\ast$\textit{-unet} \\
        \midrule
        \textbf{s2s}                    & 94.8                                 & 94.7                                                  & 97.8                                 & 39.3                & 83.1      & 91.4      & 94.0       & 75.0                   & 74.3                    & 78.6                                     & 90.8                     \\
        \textbf{u2u}                   & 92.5                                 & 99.8                                                  & 87.1                                 & 19.0                & 86.5      & 92.7      & 93.7       & 36.5                   & 74.8                    & 67.9                                     & 83.2                     \\
        \bottomrule
    \end{tabular}
    \justify
    \vspace{-0.5em}
    \emph{$\ast$-ss}: w/o second stage training,
    \emph{$k$ tgt}: $k$ target utterance(s),
    \emph{$\ast$-ca}: w/o cross-attention,
    \emph{$\ast$-lrr}: w/o learning rate reduction in the second stage training,
    \emph{$\ast$-rrc}: w/o removal of the residual connection in Extractor~1,
    \emph{$\ast$-unet}: w/o U-net-like attention.
\end{table*}

\section{Experimental setup}
\label{sec:experiments}

\subsection{Training Setup}

The whole CSTR VCTK Corpus (VCTK) \cite{veaux2016vctk} with 44 hours of speech produced by 109 speakers was used to train the FragmentVC model.
All the utterances were resampled to 16k Hz before fed into the Wav2Vec model.
The hop and window sizes for mel-spectrogram computation were chosen to ensure the length of the mel-spectrogram was identical to that of the Wav2Vec features.

FragmentVC was optimized with the AdamW optimizer \cite{loshchilov2018decoupled} (with learning rate $10^{-4}$, $\beta_1 = 0.9$, $\beta_2 = 0.999$, $\epsilon=10^{-8}$, and $weight\_decay=0.01$) for 250k steps with batch size being 16, the first 50k steps for first stage while the rest for second stage training.
In the second stage, the probability that the source utterance was included in the target utterances linearly decayed from one to zero from the 50k-th step to the 150k-th step and remained zero until the end.
We also used cosine annealing for learning rate scheduling, with 500 steps of linear warmup.
The code will be released online \footnote{\url{https://github.com/yistLin/FragmentVC}}.

\subsection{Other SOTA approaches}

For performance comparison, we took AdaIN-VC \cite{chou2019one-shot} and \textsc{AutoVC} \cite{qian2019autovc} as the SOTA any-to-any VC approaches, with officially released pretrained models also trained over VCTK, adopted in experiments.

\subsection{Vocoder}

For each model, we trained a WaveRNN-based speaker-independent vocoder \cite{lorenzo-trueba2019towards} to convert the log mel-spectrograms to waveforms.
A subset of LibriTTS \cite{zen2019libritts} (\emph{train-clean-100}) and the CMU Arctic databases (CMU) \cite{kominek2004cmu} were used to train the vocoders for 150k steps.

\subsection{Test scenarios}

Two voice conversion scenarios were evaluated: (1) seen-to-seen (s2s) for the conversion between speakers in the VCTK training dataset and (2) unseen-to-unseen (u2u) for the conversion between unseen speakers from the CMU dataset.
In both cases, we randomly sampled 1000 testing pairs within VCTK (s2s) and CMU (u2u), each including 1 utterance from a source speaker and 10 utterances from a target speaker for all models considered.
Considering that both VCTK and CMU are parallel datasets, to match the real-world scenario, the utterances with the same transcription as the source utterance were not sampled as the targets.

\subsection{Evaluation metrics}

A speaker verification (SV) system \footnote{\url{https://github.com/resemble-ai/Resemblyzer}} was adopted for objective evaluation of the converted speaker characteristics, as done in a previous work \cite{jia2018transfer}.
The SV system took a converted utterance as input and generated a fix-dimensional embedding.
The conversion was considered successful if the cosine similarity between the embeddings of the target utterance and the converted utterance exceeded a predefined threshold.
The threshold was determined based on the equal error rate (EER) of this SV system over the considered dataset.
The SV Accuracy was then the percentage of successful conversion.

For subjective evaluation of the perceptual quality, we conducted two Mean Opinion Score (MOS) tests.
In the first test, each subject was asked to listen to an authentic utterance from the target speaker and a converted result, and then to score from 1 to 5 regarding how confident they would consider these two utterances to be produced by the same speaker (5 being absolutely same and 1 absolutely different) \cite{jia2018transfer}.
In the second test, the subjects were given a converted utterance or a vocoder-reconstructed authentic utterance and asked to score from 1 to 5 how natural the utterance sounded.
For every model considered, the converted results of the same 50 randomly sampled testing pairs out of the previously used 1000 pairs were evaluated, each by at least 5 subjects.
The scores were then averaged and reported with the 95\% confidence intervals for each model.
Such subjective evaluation was conducted with the u2u scenario only, which is considered much more important than s2s for any-to-any VC.

\begin{figure*}[t]
    \centering
    \begin{subfigure}[t]{.2065\linewidth}
        \centering
        \includegraphics[width=\linewidth]{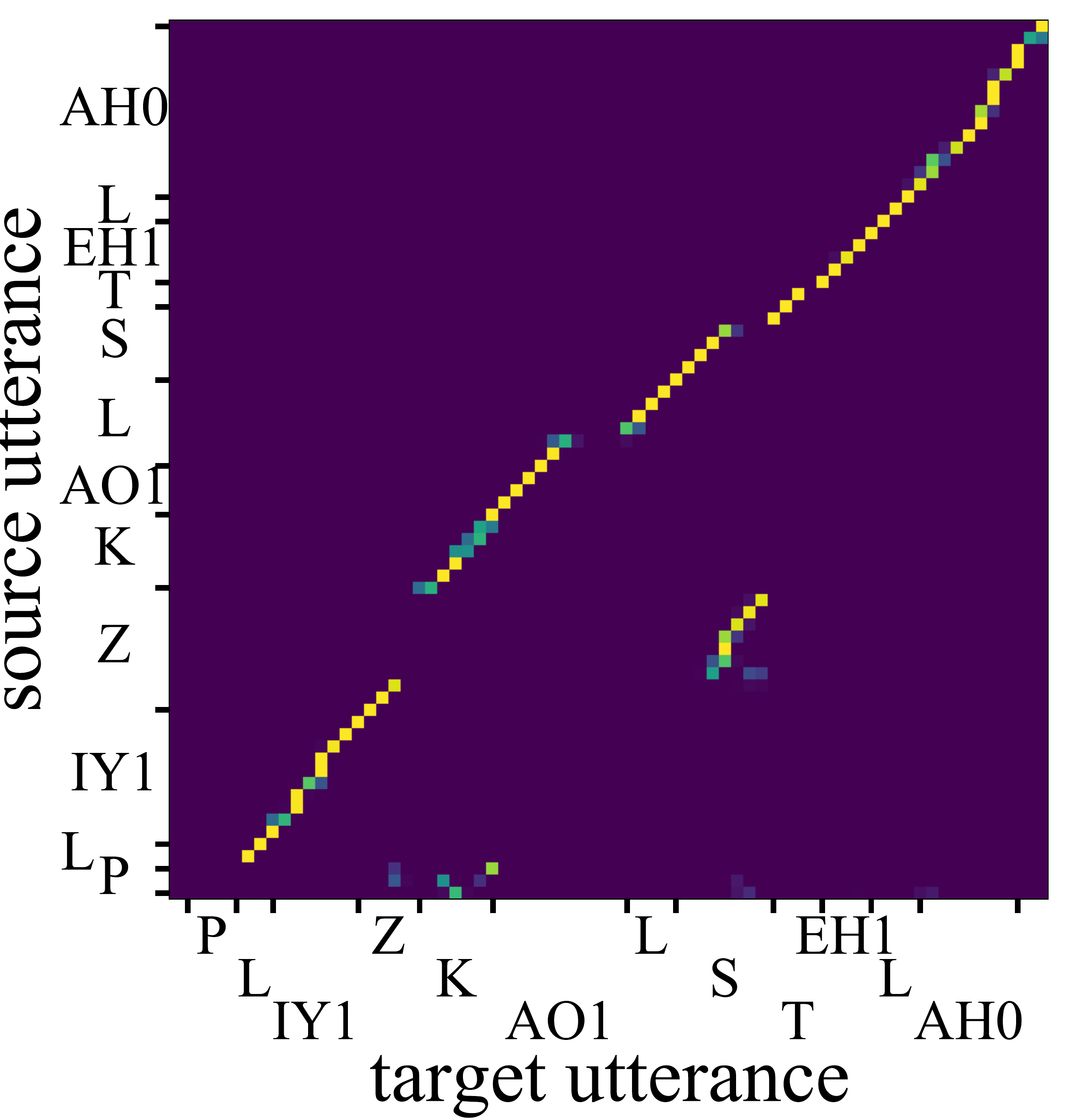}
        \vspace{-1.5em}
        \caption{Utterances of same content.}
        \label{fig:attention_same}
    \end{subfigure}
    \hfill
    \begin{subfigure}[t]{.7835\linewidth}
        \centering
        \includegraphics[width=\linewidth]{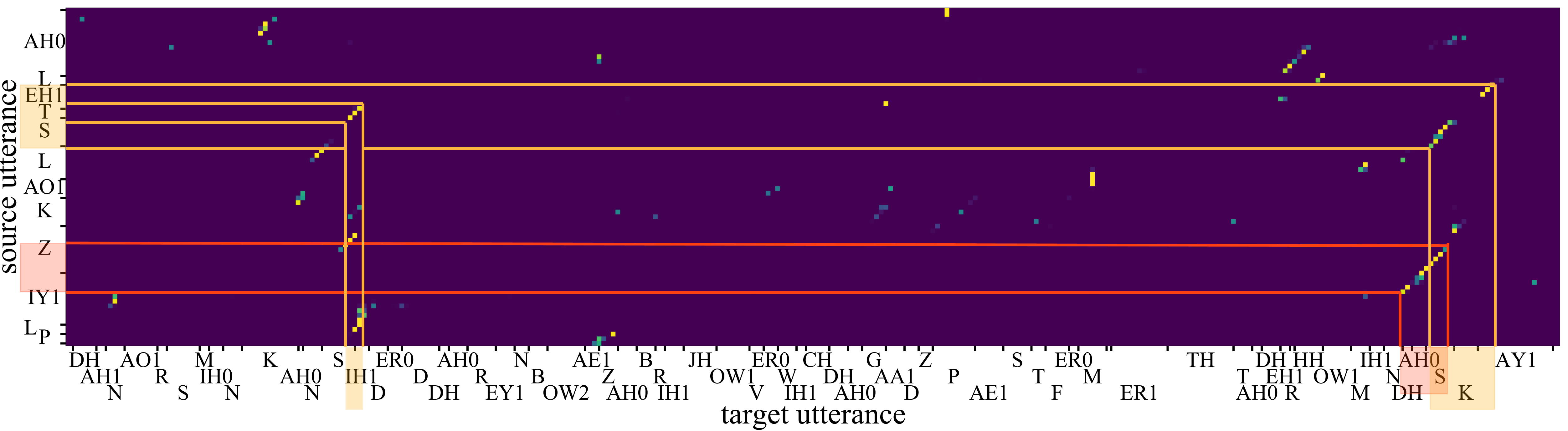}
        \vspace{-1.5em}
        \caption{Utterances of different content.}
        \label{fig:attention_diff}
    \end{subfigure}
    \vspace{-0.75em}
    \caption{
        Attention maps between Extractor 3 of the decoder and Conv1d~1 of the target encoder.
        From VCTK, utterance \emph{p225\_001} of speaker \emph{p225} as the source utterance, utterance \emph{p227\_001} (in (a)) and utterance \emph{p227\_016} (in (b)) of speaker \emph{p227} as the target utterances.
    }
    \label{fig:attention}
\end{figure*}

\section{Experimental results}
\label{sec:results}

\subsection{Performance analysis}

\begin{table}
    \setlength{\tabcolsep}{0.5pt} 
    \centering
    \caption{
        The MOS on unseen-to-unseen conversion.
        \emph{Auth.} stands for vocoder-reconstructed authentic utterances.
    }
    \label{tab:mos}
    \vspace{-0.75em}
    \begin{tabular}{cccccc}
        \toprule
        \textbf{MOS\,} &
        (a)\,$^\ast${\footnotesize Proposed}       &
        (b)\,$\ast$\textit{-ss}                    &
        (c)\,{\footnotesize AdaIN-VC}\,            &
        (d)\,{\footnotesize \textsc{AutoVC}}       &         
        (e)\,Auth.                                                              \\
        \midrule
        \textbf{Sim.}  & 3.32$\pm$0.15 & 3.81$\pm$0.15 & 2.75$\pm$0.15 & 2.12$\pm$0.14 & -- \\
        \textbf{Nat.} & 3.26$\pm$0.12 & 2.73$\pm$0.11 & 2.52$\pm$0.12 & 2.31$\pm$0.12 & 4.09$\pm$0.12 \\
        \bottomrule
    \end{tabular}
\end{table}

The results of the speaker verification accuracy are listed in Table~\ref{tab:speaker_verification_accuracies}, where columns (a) (b) are respectively for the proposed FragmentVC and that without the second stage training.
With the s2s scenario (first row), FragmentVC (with second stage training or not) achieved comparable performance in speaker characteristics conversion with AdaIN-VC (column (c)), while leaving \textsc{AutoVC} (column (d)) far behind.
As for the u2u scenario (second row), which is much more important for any-to-any VC considered, FragmentVC clearly outperformed other models.
On the other hand, column (e) (f) (g) are results of using 1, 5 or 20 utterances of the target speaker (10 used in column (a)).
It can be found the performance was clearly improved or better voice fragments can be extracted with more utterances, but even only 1 target utterance worked very well.

Table~\ref{tab:mos} lists the two MOS scores for exactly the same columns (a) (b) (c) (d) as in Table~\ref{tab:speaker_verification_accuracies}.
It can be found the proposed FragmentVC achieved significantly higher scores than the others (column (a) v.s. (c) (d)), while the vocoder-reconstructed authentic utterances (column (e)) served as the upper bound.
Column (b) of Table~\ref{tab:mos} is for the proposed FragmentVC but without the second stage training.
The second row and columns (a) (b) in Table~\ref{tab:mos} clearly verified the second stage training is very important to achieve more natural converted utterances, although at the price of slightly degrading the target speaker characteristics (column (a) v.s. (b) in the first row of Table~\ref{tab:mos} and the second row of Table~\ref{tab:speaker_verification_accuracies}).
With careful listening, we found that without the second stage training, the converted utterances sometimes sounded unsmooth, or even discontinuous, which is why the naturalness score was much lower.

\subsection{Ablation studies}

In column (h) of Table~\ref{tab:speaker_verification_accuracies}, we removed the cross-attention in the decoder and replaced the target encoder with a speaker encoder trained with GE2E loss \cite{wan2018generalized} over LibriSpeech, VCTK, and LibriTTS.
This speaker encoder generated a fixed-size speaker embedding for every target utterance, which was then averaged, linearly projected, and fed into the decoder.
The obviously inferior performance (column (h) v.s. (a)) showed the importance of the attention mechanism.

Columns (i) (j) (k) are respectively for the cases without the learning rate reduction in the second stage training, without the removal of the residual connection in Extractor~1, and without the U-Net-like architecture, namely all extractors attending on the output of Conv1d~3.
The results verified these approaches are important.

\subsection{Attention analysis}

Here we plotted and tried to analyze the attention maps between the extractors and the target encoder.
Two example attention maps between Extractor~3 of the decoder and Conv1d~1 of the target encoder are shown in Fig.~\ref{fig:attention}.
Since there are two attention heads in the extractor, we showed in these maps the root-mean-square of the attention values at each position for easier visualization.
For space limitation, only one utterance was used as the target here.
More attention plots are available on our demo page \footnote{\url{https://yistLin.github.io/FragmentVC}}.

In Fig.~\ref{fig:attention_same}, for the source and target utterance having the same content but being spoken by different speakers, a diagonal pattern can be easily seen in the attention plot, or the alignment between the two utterances was properly achieved considering their phonetic structure.
In Fig.~\ref{fig:attention_diff}, for the same source utterance but a different longer target utterance, it can be found the extractor is able to attend on phonetically similar frames in the target utterance, including finding acoustically similar voice fragments (e.g. within \texttt{/IY1 Z/} and \texttt{/AH0 S/} in the lower right corner, and \texttt{/S T EH1/} and \texttt{/S IH1/} \& \texttt{/S K AY1/} in the upper left \& right corner) for constructing the converted utterance.

\section{Conclusion}
\label{sec:conclusion}

Here we propose to achieve any-to-any voice conversion by extracting and fusing voice fragments to construct the desired utterances with attention.
The objective and subjective evaluations verified the proposed FragmentVC achieved comparable or even better performance than other SOTA approaches.
How the Wav2Vec representations actually contributed to the model, if it can be jointly learned, or if it is possible to find some other pretrained representations for the purpose are yet to be investigated.
However, we believe such attention-based approaches will be very useful for VC because of their easy implementation, flexibility and explainability.

\section{Acknowledgement}

We thank National Center for High-performance Computing (NCHC) for providing computational and storage resources.


\bibliographystyle{IEEEbib}
\bibliography{main}

\end{document}